\documentclass[preprint,aps,showpacs]{revtex4}
\usepackage{graphicx}
\usepackage{bm}
\usepackage{amsmath}

\begin{document}

\title{Stability of a Charged Particle Beam in a Resistive Plasma
 Channel\footnote{This study was supported by LDRD fund and was circulated
  as a technical report (LA-UR-93-2146) at Los Alamos.}}

\author{S. J. Han}

\affiliation{P.O. Box 4684, Los Alamos, NM 87544-4684 \\
email:sjhan@cybermesa.com}

\begin{abstract}

A self-focusing of a coasting relativistic beam in a plasma
channel that is confined by an external magnetic field is studied
as a means of reconditioning the beam emerging from a beam
injector [a radio frequency quadrupole (RFQ)] for a linac. A
detailed study of the beam stability in the self-focused beam has
been carried out. In order to explain beam filaments and the
resistive hose instability in a unified way, we treat all the
azimuthal modes in the derivation of the dispersion relation in a
finite plasma channel that exhibit many unstable modes, which are
classified by Weinberg's scheme [Steven Weinberg, J. Math. {\bf
8}, 614 (1967)]. To overcome the energy requirement of a beam
injector for a high-current, high energy linac, we suggest to add
an energy booster of a compact synchrotron to the RFQ. The
analysis is then applicable to the charged particle beam transport
in a proton accelerator, such as Large Hadron Collider (LHC) at
CERN or APT at LANL.

\end{abstract}

\maketitle

\section{\label{sec:level1} Introduction}

The application of electron beam transport in a cold plasma has
been slow, but has made great strides in recent years for a beam
focusing device \cite{Nakanishi91}, laser-guided beam transport
problems \cite{Caporaso86}, and heavy ion fusion
\cite{Boggasch91}. In these applications, the $m=1$ hose
instability, if it occurs, displaces the beam position from the
center of beam propagation. In both a beam focusing device and
heavy ion fusion, this can lead to the displacement of the focal
point, which may cause the loss of the beam to the wall of the
device. Furthermore, instabilities with higher modes ({\it e.g.,}
$m \gg 1$) can lead to beam filament.

Weinberg \cite{Weinberg64,Weinberg67} studied a relativistic beam
instability with perturbations on a particle orbit which treats
all $m$ (azimuthal quantum number) values to account for beam
deformation, and examines various types of unstable modes in the
entire frequency range.

Resistive instabilities of a charged particle beam penetrating
into a plasma channel were studied first by Longmire
\cite{Longmire}. They are complex phenomena and are still not well
understood for realistic situations. The hose ($m=1$ mode)
instability was studied by a number of  authors \cite{Longmire,
Rosenbluth60,Mjolsness63}. They have shown that a relativistic
electron beam (REB) penetrating into a cold plasma becomes
unstable to the $m=1$ mode. However, their analysis remains valid
for a background plasma whose skin depth
$(c^{2}/4\pi\sigma\omega)^{1/2}$ much larger than the beam radius.

Since then a significant progress has been made, the REB
propagating in neutral plasmas has been studied extensively by
many authors \cite{Lampe84}. Because of the mathematical
complexity of the problem, it has been necessary for many authors
to adopt simple models which are valid for specific problems, so
that the applicability of models remains limited. In general there
are serious difficulties in comparing theoretical analysis of an
instability with experiments ({\it e.g., } either a z-pinch or a
plasma produced by a gas discharge, which is then confined by a
magnetic field). We incorporate a feasible experimental condition
of interest in the theory and compare with experiment. In
particular, we are interested in the feasibility of applying the
theory of a beam-focusing ({\it i.e.,} reconditioning) in a plasma
channel, which is expected to be far more effective than the
conventional quadrupole magnet focusing in accelerator technology.
The basic idea of the technique involves the self-pinching of a
high-current of charged particles by self-magnetic fields
\cite{Bennett34}. However, the self-pinched beam is subject to
numerous instabilities associated with particle motion as
described below.

The previous analysis of beam instability was carried out with a
model of a beam penetrating into a cold plasma which is infinite
in spatial extent for mathematical convenience. However, this
approach can not be applied to the self-focusing problems for
which the background plasma must be finite. This permits one to
impose the boundary condition and thereby to obtain the dispersion
relation that gives qualitative information on the stability of
the REB.  Moreover, it does not apply to the beam focusing by a
plasmas in a drift tube confined by an external axial magnetic
field in accelerator technology, nor does it tell how the
boundaries of a finite plasma channel are formed.

A similar situation occurs in an ion-beam transport in the
accelerator technology. The excessive space-charge is the origin
of the beam divergence in a high-current, high-energy proton
linac. One important new problem arises in this connection,
however, is how do we overcome the space-charge effect in the
absence of a energy booster before injecting the beam into a
linac, which has not been addressed in the previous work. The
principal difficulty encountered in the development of the
high-current proton accelerator is that a high-current beam must
reach a critical energy to overcome the space-charge effect. This
difficulty occurs in a situation for any high-current, high-energy
proton accelerator with the direct application of a radio
frequency quadrupole (RFQ) as a beam injector
\cite{Puglisi85,Kapchinskii59,Schempp95}.

The RFQ by Kapchinski and Teplyakov \cite{Kapchinskii59} makes use
of a strong focusing with rf-electrical field, based on the same
principle as is used in a quadrupole mass spectrometer, and is
basically a homogeneous transport channel with additional
acceleration. By the geometrical modulation of quadrupole
electrodes, one generates the axial field. Thus the RFQ has a
linac structure which accelerates and focuses the beam with the
same rf fields. However, the very fact that such axial
acceleration is feasible by the geometrical modulation brings with
it new difficulties which seem to be formidable.

\begin{figure}
\includegraphics[totalheight=4in,height=4in,width=6in,
keepaspectratio=true]{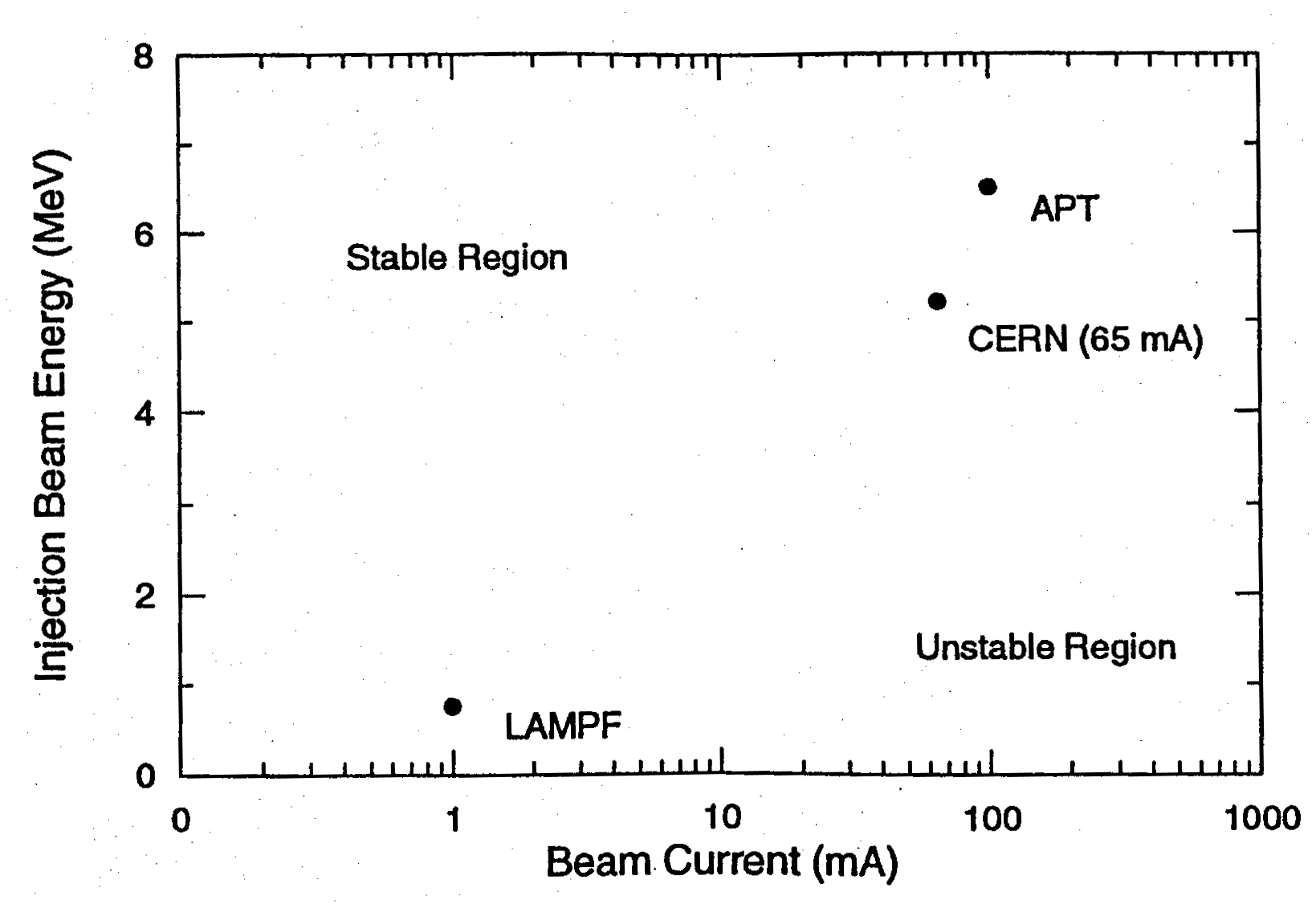}
\vspace{0.0truecm}
\caption {The stable domain, as a function of the beam
injection energy and the current, is calculated. The condition
$K^{2}-K_{1}^{2}\geq 0$ in $Eq.~(\ref{stable})$ was imposed on the
injection beam energy by $K_{1}^{2}=
\frac{2\pi e^{2}n_{0}}{c^{2}\beta^{2}M}$ for a stable particle orbit.}
\label{fig:fig1}
\end{figure}

Yet the acceleration by the RFQ is independent of beam velocity
with a large radial acceptance, which is a great advantage in the
design of a proton linac. Thus the devise offers the possibility
of utilizing it as a beam injector for a high-current linac. In
order to make use of it as an alternative to dc acceleration where
a beam injector is required, it is necessary to demonstrate the
possibility of reaching energies to about $7 Mev$ in the RFQ
before injecting the proton beam for a high-current (80mA) linac
such as Large Hadron Collider (LHC) at CERN as shown in Figure
\ref{fig:fig1}. This prerequisite is determined by solving the
equations of motion in the presence of self-field in a quadrupole
structure in a linac \cite{Courant52}. The energy requirements
impose additional restrictions on the feasibility of the LHC at
CERN.

The requirements have been appreciated for some time now because
of our inability to invent a new device that overcome the beam
energy requirement for a beam injector for a high-current,
high-energy accelerator; however it was not difficult to overcome
the limitation for a low-current accelerator such as the proton
accelerator at Los Alamos (LAMPF).

From a theoretical point of view, the reason for this limitation
can be understood, to a great extent, with the observations that
the beam focusing force by a quadrupole magnet in a linac is
globally a second-order effect, and the geometrical requirements
of a high focusing and a large magnetic aperture cannot be
achieved simultaneously for a diverging high-current proton beam
\cite{Courant52,note1}.

There are several review papers on the progress of RFQ
\cite{Schempp95} and its wide application in many laboratories.
The maximum attainable beam energy still remains about the same in
spite of intense effort to overcome the limitation posed by a
high-current, high-energy accelerator.

For example, we give here the numbers for a figure of merits for
the LHC planned at CERN \cite{Alessi89}; with $100 mA$ input to
the RFQ, $80 mA$ proton beam was accelerated to $520 KeV$ by RFQ,
of which $65 mA$ proton beam was accelerated through the first
linac tank. Subsequent linac tank cannot accelerate the high
current beam and loses a significant amount of the beam as shown
in Fig. 1 which was obtained by solving Eqs. 13 a-b of Courant,
Livingston and Snyder \cite{Courant52}. It is this fundamental
limitation of RFQ that makes it exceedingly difficult to build a
high-current, high-energy accelerator such as LHC at CERN. In
general, the higher the current density of a proton beam, the
higher the injection energy will be in order for quadrupole
magnets in a proton linac to transport the beam with a sufficient
focusing force.

The calculation of the above energy requirement is most simply
carried out by introducing the condition that the particles remain
in oscillating orbits in one direction in the presence of
self-fields and quadrupole magnetic fields with a given field
gradient:
\begin{equation}
\frac{d^{2}y}{dx^{2}}+(K^{2}-K_{1}^{2})y=0, \label{stable}
\end{equation}
where $(K^{2}-K_{1}^{2})\geq 0$, $K^{2}=(dB_{z}/dy)/(BR)$,
$K_{1}^{2}=\frac{2\pi^{2}}{c^{2}\beta^{2}}\frac{n_{0}}{M}$ with
$\gamma\cong 1.0$, $dB_{z}/dy=3.94\times10^{3} Gauss/cm$
\cite{Courant52}, and $P_{\bot}=3.0\times 10^{-4}BR (Gauss-cm)$
\cite{Jackson75}.

The Figure \ref{fig:fig1} shows the critical beam injection energy
for a given current density, and shows the domain of beam
stability. The occurrence of beam divergence due to the space
charge effect is the origin of the beam divergence which results
from our attempt to accelerate a high-current beam in a linac,
since the focusing force by a quadrupole magnet is not
sufficiently strong to overcome the space-charge, which poses the
fundamental limitation of the current accelerator technology
\cite{Schempp95,Picardi91}.

Although the demonstration of successful operation of RFQ in
low-current proton accelerator at Los Alamos (LAMPF) has raised a
hope of developing a high-energy, high-current proton accelerator,
the fundamental difficulty in the development of a high-energy,
high-current accelerator is remains unsolved; that the
high-current beam diverges (bursts) as it emerges from RFQ in a
linac is a real challenge to overcome. This difficulty occurs even
if the best possible design parameters of RFQ to overcome the
Coulomb repulsion in the beam are employed \cite{Schempp95}. Yet
the problem of space-charge of a high-current beam was the
original motivation of Kapchinski and Teplyakov's concept of RFQ
as a possible beam injector for a linac.

One important new possibility that arises in this connection,
however, is a utilization of self-focusing by a neutral plasma
channel which has not been treated in the previous papers. The
existence of the current limitation in the RFQ has made it
necessary to study the possibility of space-charge compensated
beam transport \cite{Nezlin68} in the development of a
high-current, high-energy proton accelerator such as a large
hadron collider (LHC) at CERN.

An important question one should ask: is there any method that can
be applied to reconditioning the perturbed beam emerging from RFQ
before injecting it to a linac? A well-known application of
space-charge compensated beam \cite{Nezlin68} is an effective,
useful concept to overcome the problem of space-charge effect
which is the major stumbling block in a high-current beam
transport. The decisive advantage of this approach was
demonstrated in a high-current electron accelerator by shielding
the space-charge with a quiescent plasma \cite{Nakanishi91}. The
focusing force may reach to the value, that is greater than that
of a super-conducting magnet by several order. Here we assume the
presence of an over-dense plasma in which the plasma density is
much higher than that of a beam. Thus the Coulomb repulsive force
due to space-charge in the beam is balanced by the self-field of a
bunched beam maintaining a constant radius. However, the
effectiveness of  the space-charge neutralized beam transport
depends on $\beta=v/c$. Consequently, since the beam emerging from
RFQ is in non-relativistic domain ($\beta=v/c\ll$), the
feasibility of a beam reconditioning is out of the question. This
technique can be applied only to a charged particle beam with the
velocity $\beta=v/c \geq 1$ by the combined use of the synchrotron
as the energy booster.

For some time it has been realized that it might not be possible
to make use of the RFQ  as a beam injector for a high-current
high-energy linac \cite{Alessi89,Schempp95}. The simplest and
probably the best, way of accelerating the high-current proton
beam is to apply the phase-locking method in the synchro-cyclotron
\cite{Bohm46,Bohm47}, but it will be difficult to efficiently
extract the high-density beam. Yet it seemed feasible to attempt
the experiment with a combination of a synchrotron with the RFQ ,
instead of the betatron injection \cite{Pollock46}, to accelerate
the beam to the relativistic domain, in spite of the fact that the
beam is being lost in the extraction process. Thus the role of the
RFQ is promising as a pre-injector for a high-current, high-energy
accelerator provided that the beam emerging from the RFQ can be
reconditioned by a plasma channel described below. In PHERMEX
facility at Los Alamos, the high-energy $\gamma$-rays are produced
by short pulsed high-energy electron beam. To make a uniformly
diverging x-ray source, it is essential to have a highly focused
beam, a pencil beam, that can be achieved only by a plasma
focusing.

Finally, the dependence of $\beta=v/c$ on the effectiveness of
space-charge compensated beam transport arises from the form of
self magnetic filed which is the primary focusing force. Since the
stability due to beam modulation has been studied in a plasma
channel \cite{Weinberg64} and a somewhat more detailed study of
the model has been already made by Weinberg himself
\cite{Weinberg67}, here we study the beam reconditioning with the
assumption that the beam is accelerated {\it somehow} to the
relativistic domain in which there is no beam modulation and is
coasting in a plasma channel confined by an {\it external magnetic
field}. In this process the beam is being reconditioned and is to
be transferred to a linac in the next stage of acceleration
\cite{Alessi89}. As we shall see, this simple model calculation
presents a rather formidable mathematical challenge which defines
the ultimate fate of LHC at CERN.

We shall be interested here primarily in the stability of the
relativistic beam, in particular, the hose instability of a
coasting relativistic beam in a plasma channel confined by the
external magnetic field since this is the one which may lead to
the loss of charged particles before the beam is injected to the
main linac. An elegant formalism of beam instability is provided
by the use of the Lagrangian displacement vectors which have been
introduced by Bernstein, Frieman, Kruskal and Kulsrud in their
study of a Rayleigh type energy principle for hydro-magnetic
stability problems \cite{BFKK58} and is applied to the
Boltzman-Vlasov equation for plasmas by Low \cite{Low58}. The
treatment given by these authors has shown that the method is
indeed very powerful in treating stability problems. Here we make
use of the concept of Lagrangian displacement vectors
\cite{BFKK58} to give an alternative, more mathematically elegant
method leading to Weinberg's results \cite{Weinberg67} in the
limit of vanishing external magnetic field \cite{comments1}.

The present model is essentially equivalent to that of Weinberg
\cite{Weinberg67}, but is different in one major respect, that is,
we incorporate the external magnetic field to confine the
necessary plasma inside the drift tube, so that we may impose
proper boundary conditions. The space-charge compensated beam
transport is essentially the same for both the high-current
electron rf linac (ILC) and the high-current proton linac (LHC),
although it is far more useful to overcome the difficulty of
focusing the beam in a proton linac.

With these qualitative remarks as an introduction, we proceed to
the development of the space-charge compensated beam transport in
linac. We therefore focus on the stability of the REB penetrating
into a plasma channel confined by a weak external axial magnetic
field ({\it i.e.,} $B_{0}^{ext} \cong 1.0 KG$), consistent with
the experiment \cite{Nakanishi91}. Since we introduce
perturbations in particle orbits which are determined by the
external magnetic field and the self-field, it is important to
include all the relevant fields in the equations of motion of a
particle.

It is the purpose of this paper to present the essence of our
results and some of the details of how the dispersion relation is
obtained. The present work is a generalization of Weinberg's work
on the general resistive instability \cite{Weinberg67}. To make
the discussion of our model clearer, we derive the dispersion
relation by Weinberg \cite{Weinberg67} by taking the limit
$\omega_{c}=0$ which implies that no external magnetic field
applied to the system. This will illustrate the versatility of the
formal application of the Lagrangian displacement vectors in the
study of beam instability. It is shown here that, even in the
presence of an axial guide magnetic field, the REB remains
unstable to various perturbations. These include the $m=1$ hose
instability and the set of the higher modes that are unstable to
various perturbations (type A, B, ${\cal A}$, ${\cal B}$, C, D,
and ${\cal D}$).

\section{\label{sec:level1}Basic Equations}

There are two avenues of approach to this problem. One is to
pursue the approach of Weinberg \cite{Weinberg67} and consider the
the first-order effects by the external magnetic field. Though
some details of his theory are open to criticism, there is no
doubt that the paper points to the right direction for this
difficult problem. However I must confess that I have been unable
to follow the early part, "Introduction", of his paper. A brief
study of Weinberg's paper has convinced me to take a more formal
approach that is transparent. Indeed it is much more preferable
mathematically to apply a straightforward technique to a complex
perturbation problem of a many-body system. Moreover, without a
confining external magnetic field, Weinberg's model does not allow
one to impose the necessary boundary conditions that are essential
to derive his dispersion relations, Eq. 1-9 \cite{Weinberg67} and
is therefore inconsistent with his object of the problem that he
set out to solve. Hence the object of this paper is also to
establish what is correct in Weinberg's analysis, and to remove
some of ambiguities.

To obtain a proper dispersion relation, it is first necessary to
specify the unperturbed motion of an electron in the beam. The REB
is assumed to move in the $z$-direction with the average velocity
$\bm v$. The self-field is then given by
\begin{equation}
B_{s}=\frac{4\pi e
\beta}{r}\int_{0}^{r}r^{\prime}n(r^{\prime})dr^{\prime},
\label{self1}
\end{equation}
where $n(r)$ is the particle density of the REB.

Then the pinch effect can be described by the force (MHD) equation
\begin{equation}
\label{pinch}
\frac{dp}{dr}=\frac{d}{dr}\left(\frac{B^{2}_{s}}{8\pi}\right)+
             \frac{B^{2}_{s}}{4\pi r},
\end{equation}
where $p$ is the hydrodynamic pressure on the beam. The solution
of Eq.~(\ref{pinch}) is then
\begin{equation}
\label{pressure}
p(r)=p_{0}-\frac{1}{8\pi}\int_{0}^{r}\frac{1}{r^{2}}\frac{d}{dr}(r^{2}B_{s}^{2})dr
\end{equation}
with the axial pressure $p_{0}$
\begin{equation}
p_{0}=\frac{1}{8\pi}\int_{0}^{R}\frac{1}{r^{2}}\frac{d}{dr}(r^{2}B_{s}^{2})dr.
\end{equation}
Hence Eq.~(\ref{pressure}) can be written
\begin{equation}
p(r)=\frac{1}{8\pi}\int_{r}^{R}\frac{1}{r^{2}}\frac{d}{dr}(r^{2}B_{s}^{2})dr.
\end{equation}

In addition we define
\begin{equation}
\alpha^{2}\equiv\frac{4\pi e^{2}\beta^{2}}{\gamma
M}\frac{1}{r^{2}}
           \int_{0}^{r}r^{\prime}n(r^{\prime})dr^{\prime},
\end{equation}
which is equal to
$\alpha^{2}\equiv\omega_{\beta}^{2}=e\beta/(\gamma M)(B_{s}/r)$,
where $(r^{2}\alpha^{2})^{\prime}=(\beta^{2}r)\omega^{2}_{p}$ and
$\omega^{2}_{p}=4\pi e^{2}n /(\gamma M)$.

The rotational motion of a coasting beam can be described by the
equation of motion in the presence of an external and a self
magnetic fields:
\begin{equation}
\label{Zero} (\frac{\partial}{\partial t}+\bm v \cdot\bm\nabla)
(v_{\theta}{\hat e}_{\theta}+ v_{z}{\hat e}_{z})= \frac{e}{\gamma
M}(\bm\beta\times \bm B^{ext}+\bm\beta\times\bm B_{s}),
\end{equation}
where $\omega_{c}=eB^{ext}/(\gamma Mc)=ecB^{ext}/{\cal E}$ and
$v_{z}=constant$.

One can easily show from Eq.~(\ref{Zero}) that the angular
frequencies are given by
\begin{equation}
\label{angular}
\omega_{\pm}=\frac{1}{2}[-\omega_{c}\pm\sqrt{\omega_{c}^{2}+4\alpha^{2}}].
\end{equation}

Hence the transverse motion of a particle in the polar coordinates
can be described by the equation:
\begin{equation}
\ddot{\bm r}=\dot{\bm r}\times\bm{\omega}_{c}-\alpha^{2}\bm {r},
\end{equation}
where $\omega_{c}=eB^{ext}/(\gamma M c)$. Here $\gamma$ and $M$
are the Lorentz factor and the mass of a charged particle
respectively.

If we now define the rotational velocities as
$\bm{u}_{\pm}=\dot{\bm r}-\omega_{\mp}\bm{e}_{z}\times\bm{r}$,
then we can show that
$(\omega_{+}-\omega_{-})\bm{r}=-\bm{e}_{z}\times(\bm{u}_{+}-\bm{u}_{-})$,
which describes a uniform circular motion of an electron with
frequencies $\omega_{\pm}$ in the $r-\theta$ plane. Furthermore,
by taking the Landau gauge
$\bm{A}=\frac{1}{2}\bm{B}^{ext}_{0}\times\bm{r}$, the canonical
momentum becomes $\bm{p}=\gamma M(\bm{u}_{+}+\bm{u}_{-})/2$. The
rotational velocities are not a constant of motion; but the
canonical momentum $\bm{p}$ can be related to the Hamiltonian of
the system \cite{Lippmann49} and hence we employ $\bm{p}$ in the
derivation of a dispersion relation.

If a beam is in a relativistic motion in the presence of electric
and magnetic fields, then its translational motion of a particle
in a fluid element must be in a manifestly covariant form
\cite{Jackson75}:
\begin{equation}
\label{covariant} \frac{d\bm\beta}{dt}=\frac{e}{\gamma Mc}[\bm
E+\bm\beta\times\bm B
                    -\bm\beta(\bm\beta\cdot\bm E)],
\end{equation}
where $\bm\beta=\bm v/c$.

Next for a uniform beam with density profile,
\begin{equation}
\label{density} n(r)=\begin{cases}
     n & \text{if $r<a$}\\
     0 & \text{if $r>a$},
     \end{cases}
\end{equation}

the rotation frequency is given by
 \begin{equation}
 \label{alpha}
 \alpha^{2}(r)=\begin{cases}
          \omega_{\beta}^{2}& \text{if $r<a$},\\
          \omega_{\beta}^{2}a^{2}/r^{2}& \text{if $r>a$}.
          \end{cases}
 \end{equation}
We often simply write for a uniform beam
$\alpha^{2}\equiv\omega_{\beta}^{2}= 2\pi e^{2}n\beta^{2}/\gamma
M$ if there is no ambiguity.

\section{\label{sec:level1} First-Order Equations}

The treatment of the first-order equations of motion in Weinberg's
paper did not appear to be in the most convenient form which is
difficult to follow through \cite{Weinberg67}. In fact it is
almost impossible to introduce an external magnetic field in a
formal analysis in Weinberg's approach, which is necessary for a
systematic study of the beam instabilities in a finite plasma
channel.

\subsection{\label{sec:level2}Calculation of Perturbed Fields}

In this section, first we write down the first-order field
equations from the Maxwell equations assuming that all variables
vary as $f_{i}(r)\exp(im\theta+ikz-i\omega t)$:
\begin{subequations}
\label{Maxwell1}
 \begin{eqnarray}
q^{2}E_{1r} &=& -\frac{m}{r}\frac{\omega}{c}B_{1z}+
ik\frac{\partial
E_{1z}}{\partial r}-4\pi i\frac{\omega}{c^{2}}J_{1r},\\
q^{2}E_{1\theta} &=& -\frac{m}{r}k
E_{1z}-i\frac{\omega}{c}\frac{\partial B_{1z}}
{\partial r}-4\pi i\frac{\omega}{c^{2}}J_{1\theta},\\
q^{2}B_{1\theta} &=& i\frac{\omega}{c}\frac{\partial
E_{1z}}{\partial r}
-k\frac{m}{r}B_{1z}-4\pi i\frac{k}{c}J_{1r},\\
q^{2}B_{1r} &=& \frac{m}{r}\frac{\omega}{c} E_{1z}
+ik\frac{\partial B_{1z}}{\partial r}+4\pi
i\frac{k}{c}J_{1\theta},
 \end{eqnarray}
 \end{subequations}
where $q^{2}=\omega^{2}/c^{2}-k^{2}$.

Second from Eqs.~(\ref{Maxwell1}) we obtain the decoupled field
equations:
\begin{subequations}
\begin{eqnarray}
\label{Electric1} \frac{1}{r}\frac{\partial}{\partial r}
\left(r\frac{\partial E_{1z}}{\partial
r}\right)-\frac{m^{2}}{r^{2}}E_{1z} +q^{2}E_{1z} &=
&\frac{4\pi\omega}{c^{2}(k^{2}+q^{2})}\left[-iq^{2}J_{1z}
+i\frac{m}{r}kJ_{1\theta}+
\frac{k}{r}\frac{\partial}{\partial r}(rJ_{1r})\right]\\
\nonumber\\
\label{Solenoid1} \frac{1}{r}\frac{\partial}{\partial r}
\left(r\frac{\partial B_{1z}}{\partial
r}\right)-\frac{m^{2}}{r^{2}}B_{1z} +q^{2}B_{1z}
&=&-\frac{4\pi}{rc}\left[-imJ_{1r}+ \frac{\partial}{\partial
r}(rJ_{1\theta})\right].
\end{eqnarray}
\end{subequations}

Here we shall examine in detail the first-order equations of
motion by introducing the displacement vector in a particle orbit
$\bm{\xi}(\bm{r}_{0},t)$, where $\bm{r}_{0}$ describes the
unperturbed trajectory of a charged particle (an electron or a
proton) and $t$ is the time. Here the displacement $\bm{\xi}$ is
defined by the equation $\bm{r}=\bm{r}_{0} +
\bm{\xi}(\bm{r}_{0},t)$.  Upon introducing the Lagrangian
displacement vector $\bm{\xi}$, it is possible to expand the
velocity in terms of $\bm{\xi}$. This makes it simple to derive
the first-order equations of motion in the presence of an external
magnetic field and the perturbed fields. In particular, we must
modify the equation of motion for a fluid element that moves with
relativistic speed in the presence of electric and magnetic fields
as described in Eq.~(\ref{covariant}).

Next we expand the velocity to the first-order in $\bm{\xi}$
defined by $\bm r=\bm r_{0}+\bm{\xi}$, limiting to the fast mode
for the time being,
\begin{equation}
\label{velocity1} \bm{v}(\bm{r}_{0}+\bm{\xi})=\bm{v}_{0}+ \bm
v_{1}(\bm\xi),
\end{equation}
where $\bm v_{1}$ is given by the following equation.
\begin{eqnarray}
\bm v_{1} &= &\frac{\partial \bm \xi}{\partial t}
                       + \bm v_{0}\cdot\bm\nabla\bm\xi,\nonumber\\
          & = &-i(\Omega-m\omega_{+})\bm\xi+
          \omega_{+}(\xi_{r}\hat{e}_{\theta}-\xi_{\theta}\hat{e}_{r}),
\end{eqnarray}
where we have limited to the fast mode $\omega_{+}$ in
Eq.~(\ref{angular}) and will repeat the same calculations later
for the slow mode $\omega_{-}$ to check algebras.

We then derive the first-order equation of motion from
Eqs.~(\ref{covariant}) and (\ref{Maxwell1}) after a short algebra:
\begin{eqnarray}
\label{First} \frac{\partial^{2}\bm{\xi}}{\partial
t^{2}}+2\bm{v}_{0}\cdot{\bm\nabla}\frac{\partial\bm{\xi}}{\partial
t} +
(\bm{v}_{0}\cdot{\bm\nabla})(\bm{v}_{0}\cdot{\bm\nabla})\bm{\xi}=\frac{e}{\gamma
M}
[\bm{E}_{1}+\bm{\beta}\times\bm{B}_{1}+\bm{\beta}_{1}\times\bm{B}^{ext}_{0}\\\nonumber
+\bm{\beta}_{1}\times\bm{B}^{s}_{0}+
\bm{\beta}\times(\bm{\xi}\cdot{\bm\nabla})\bm{B}^{s}_{0}
 -\bm\beta(\bm\beta\cdot\bm E_{1})],
\end{eqnarray}
where $\bm{\beta}=\bm v/c$ and the subscript $1$ denotes the
first-order.

\subsection{\label{sec:level2}Calculation of Perturbed Density}

To calculate the perturbed current density, we must still find a
way of expressing the perturbed beam density in terms of
displacement vectors. A simple alternative method to that of
Weinberg \cite{Weinberg67} is to linearize the equation of
continuity:
\begin{equation}
\label{cont} \frac{\partial n}{\partial t}+\bm\nabla(n\bm v)=0,
\end{equation}
with care on $\bm x\rightarrow \bm x+\bm\xi$ and $n(\bm
x+\bm\xi,t)= n_{0}+(\bm\xi\cdot\bm\nabla)n_{0}+n_{1}$ and then to
pick up the first-order terms in the expansion of
Eq.~(\ref{cont}):
\begin{equation}
\label{rho1} \frac{\partial n_{1}}{\partial
t}+\bm\nabla\cdot(n_{0}\bm v_{1})
      +\bm\nabla\cdot(n_{1}\bm v_{0})+
      (\bm v_{0}\cdot\bm\nabla)(\bm\xi\cdot\bm\nabla)n_{0}
      -\bm\nabla(\bm\xi\cdot\bm\nabla)\cdot(n_{0}\bm v_{0})=0,
\end{equation}
where we have used $\nabla\rightarrow
\nabla_{0}-\nabla_{0}\,\bm{\xi}\cdot\nabla_{0}$ in the expansion
and then dropped the subscript from $\bm\nabla_{0}$.

Returning to Eq.~(\ref{rho1}) with
$\xi_{i}=\xi_{i}(r)e^{i(kz+m\theta-\omega t)}$, and expressing
$\xi_{r}$ and $\xi_{\theta}$, we obtain
\begin{equation}
 -i(\Omega-m\omega_{+})n_{1}-
i(\Omega-m\omega_{+})\bm\nabla\cdot(n_{0}\bm\xi)=0,
\end{equation}
which yields the perturbed density
$n_{1}=-\bm\nabla\cdot(n_{0}\bm\xi)$. This is exactly the same as
Weinberg's derivation of the perturbed density which is much more
elegant a method of deriving the perturbed density (Eq.~(5.10) of
Weinberg's \cite{Weinberg67}).

Hence we write the density perturbation as
\begin{equation}
\label{density1} n(\bm x+\bm\xi)=n_{0}(1-\bm\nabla\cdot\bm\xi),
\end{equation}
for a uniform beam.

This is perhaps the most crucial equation that describes the
collective effects. We can impose an appropriate jump condition at
the beam-plasma boundary with the equation and derive the
first-order equations of motion in terms of the displacement
vector $\bm{\xi}$ in the many-particle system. As mentioned
earlier, there is a domain of parameters in which we can find a
tractable solution for the hose instability in a plasma channel,
for which we refer the reader to Weinberg' paper
\cite{Weinberg67}.

\section{\label{sec:level1} Dispersion Relations for A Uniform Beam}

Since the hose instability in a modulated beam that includes the
beam bunching in the non-relativistic domain is exceedingly
complicated, we limit the hose instability in the relativistic
domain in which a model illustrates a charge-compensated beam
transport consistent with the experiment \cite{Nakanishi91}.
Moreover, we set aside the question of whether a reconditioning of
a relativistic beam emerging from a beam injector can be studied
quantitatively with the aid of the coasting beam model in a plasma
channel. Hence we leave the question of an appropriate beam
injector open for future discussion.

To obtain the desired dispersion relation for a coasting beam for
which wake fields are negligible, we assume the first-order
quantities vary as $f_{i}e^{-i(\omega t-m\theta-kz)}$ as before.
Here we limit our calculation to the fast mode $\omega_{+}$. Then
Eq.~(\ref{First}) yields after a brief algebra,
\begin{subequations}
\label{FirstD} \label{allequations}
\begin{eqnarray}
\label{Firsta}
-[(\Omega-m\omega_{+})^{2}-r(\alpha^{2})^{\prime}]\xi_{r}
+i(\Omega-m\omega_{+})(2\omega_{+}+\omega_{c})\xi_{\theta}
&=&\frac{e}{\gamma M}(E_{1r}-\beta B_{1\theta}),\\
\label{Firstb}
-i(\Omega-m\omega_{+})(2\omega_{+}+\omega_{c})\xi_{r}-
(\Omega-m\omega_{+})^{2}\xi_{\theta}
& =&\frac{e}{\gamma M}(E_{1\theta}+\beta B_{1r}),\\
\label{Firstc}
 -(\Omega-m\omega_{+})^{2}\xi_{z}
&=&\frac{e}{\gamma^{3}M}E_{1z},
\end{eqnarray}
\end{subequations}
where $\Omega=\omega-kv$.

The significance of the right hand-side of Eq.~(\ref{Firstc}) is
easily seen if we compare it with those of Eq.~(\ref{Firsta}) and
Eq.~(\ref{Firstb}). In the relativistic domain for which
$\gamma\gg1$, Eq.~(\ref{First}) is essentially reduced to a
two-dimensional ($r$ and $\theta$) problem, since $\xi_{z}$ is too
small to retain and may be dropped in the analysis. Hence there is
no beam bunching in the relativistic domain. This observation is
consistent with the order-analysis of Weinberg
\cite{Weinberg67,Weinberg64} and gives a hint that there is a
domain in which we may find a tractable solution for the beam
instability. The straightforward solution of these equations is
still very difficult although perhaps not impossible. Hence we
follow Weinberg's analysis very closely to obtain approximate
solutions in the limiting cases.

With the same set of basic assumptions as in Weinberg's analysis
$|\omega|a\ll 1$, $|k|a\ll 1$, and $|q|a \approx 1$, we look for
the parameters of the beam and plasma $c^{2}/(4\pi\sigma v_{0})\ll
1$ and $\omega_{\beta}a/v_{0}\ll 1$ such that $\omega a\ll v_{0}$
and $|k|a\ll 1$. Repeating the same order analysis of $J_{1r}$,
$J_{1\theta}$ and $J_{1z}$ in Eq.~(\ref{Electric1}), we arrive at
the equation:
\begin{equation}
\frac{1}{r}\frac{\partial}{\partial r} \left(r\frac{\partial
E_{1z}}{\partial r}\right)-\frac{m^{2}}{r^{2}}E_{1z} +q^{2}E_{1z}
= \frac{-4\pi ie \beta\omega}{c}n_{1}
\end{equation}

In the following analysis we follow closely the procedure of
Weinberg \cite{Weinberg67} and make the same basic assumptions,
$|\omega|a\ll v$, $|k|a\ll 1$, and $|q|a\sim 1$, where $q^{2}=
-k^{2}+(i\omega/c^{2})(4\pi \sigma-i\omega)\rightarrow 4\pi
i\sigma\omega/c^{2}$. Here $a$ is the beam radius.

By trivial extension of the arguments leading to Eq. (3.30) of
Weinberg \cite{Weinberg67}, we obtain

\begin{equation}
\label{wave1} \frac{1}{r} \frac{d}{dr}r \frac{d}{dr} {\cal E}
 -\frac{m^{2}}{r^{2}}{\cal E}+q^{2}{\cal E} = \frac{-4\pi
iev\omega}{c^{2}}n_{1},
\end{equation}
where $E_{1z}={\cal E}(r)e^{-i\omega t+ikz+im \theta}$. Here we
have taken the plasma current $\bm{J}_{p}=\sigma\bm{E}_{1}$, where
$\sigma$ is a scalar. In a dense plasma, Ohm's law in its simple
form remains valid in a wide frequency range. Hence the Hall
effect in the REB is negligible. This assumption of scalar
conductivity in our model is reasonable, because
$\omega_{c}/\nu_{ei}\ll 1$, where $\omega_{c}\sim 5.0\times10^{8}
sec^{-1}$ and $\nu_{ei}\sim 2.2\times10^{9}sec^{-1}$. Here
$\sigma\sim 1.4\times 10^{12}sec^{-1}$ for $n_{p}\sim
10^{12}/cm^{3}$ at $kT=12eV$ \cite{Jackson75}.

We rewrite Eqs.~(\ref{FirstD}) to calculate  $n_{1}$
\begin{subequations}
\label{allequations} \label{First2}
\begin{eqnarray}
\label{Firsta2}
-[(\Omega-m\omega_{+})^{2}-r(\alpha^{2})^{\prime}]\xi_{r}
+i(\Omega-m\omega_{+})(2\omega_{+}+\omega_{c})\xi_{\theta}
&=& \frac{{\cal F}_{r}(r)}{\gamma M}\\
\label{Firstb2}
+i(\Omega-m\omega_{+})(2\omega_{+}+\omega_{c})\xi_{r}+
(\Omega-m\omega_{+})^{2}\xi_{\theta} & =& \frac{{\cal
F}_{\theta}(r)}{\gamma M},
\end{eqnarray}
\end{subequations}
where $\Omega=\omega-kv$ and $(\alpha^{2})^{\prime}=0$ in
Eq.~(\ref{Firsta2}) for a uniform beam.

Here we have extended Weinberg's analysis on the right-hand sides
of Eq.~(\ref{Firsta}) and Eq.~(\ref{Firstb}) to write
\begin{subequations}
\label{force1}
\begin{eqnarray}
\label{force1r}
 F_{1r}&= &\frac{e}{\gamma M}(E_{1r}-\beta B_{1\theta})=
 -\frac{iev}{\gamma M\omega}E_{1z}^{\prime}=
 \exp{[-i(\omega t-m\theta-ikz)]}\frac{{\cal F}_{r}(r)}{\gamma M},\\
\label{force1th} F_{1\theta}&= &\frac{e}{\gamma
M}(E_{1\theta}+\beta B_{1r})= \frac{ev}{\gamma M}E_{1z}=
\exp{[-i(\omega t-m\theta-ikz)]}\frac{{\cal F}_{\theta}(r)}{\gamma
M},
\end{eqnarray}
\end{subequations}
where we take approximate values of $F_{i}$ in terms of the
electric field $E_{1z}$ \cite{Weinberg67}. That is,
\begin{subequations}
\label{Force}
\begin{eqnarray}
{\cal F}_{r}(r) &= &\frac{-ie v}{\omega}{\cal E}(r)^{\prime}, \\
{\cal F}_{\theta}(r) &= &\frac{e v m}{\omega}\frac{{\cal
E}(r)}{r}.
\end{eqnarray}
\end{subequations}

Solving for $\bm \xi$ in Eq.~(\ref{First2}) in terms of the
perturbed field $\bm E_{1z}= e^{-i\omega t-im\theta +ikz}{\cal
E}$, we have
\begin{subequations}
\label{displacements}
\begin{eqnarray}
\label{radial1} \xi_{r} &= &\frac{2iev}{\gamma
M\omega}\frac{1}{\Xi(\omega)} {\left[{\cal E}^{\prime}(r) -
\frac{(2\omega_{+}+\omega_{c})}{(\Omega-m\omega_{+})}
\left(\frac{m}{r}\right){\cal E}(r)\right]}\\
\label{theta1} \xi_{\theta} &= &-\frac{ev}{\gamma
M\omega}\frac{1}{\Xi(\omega)}
\left[-\frac{(2\omega_{+}+\omega_{c})} {(\Omega-m\omega_{+})}{\cal
E}^{\prime}(r)+ [1-\frac{r(\alpha^{2})^{\prime}}{(\Omega-
m\omega_{+})^{2}}]\left(\frac{m}{r}\right){\cal E}(r)\right],
\end{eqnarray}
\end{subequations}
where $\Xi(\omega)=(\Omega-m\omega_{+})^{2}-
4\alpha^{2}-r(\alpha)^{\prime}-\omega_{c}^{2}$.

Making use of the identities
$n(r)=(r^{2}\alpha^{2})^{\prime}(\gamma M)/(4\pi e^{2}\beta^{2}r)$
and $2\omega_{+}+\omega_{c}=\sqrt{\omega_{c}^{2}+\alpha^{2}}$ and
defining $n_{1}=e^{i(m\theta+kz-\omega)}{\cal
N}(r)=-\bm\nabla\cdot(n\bm\xi)$,  we obtain the perturbed density
in terms of $\cal {E}$ as
\begin{equation}
{\cal N}(r)=\frac{ic^{2}}{4\pi
ev\omega}\left[\frac{1}{r}\frac{d}{dr}\left(rf_{+}\frac{{\cal
E}}{dr}\right)-\frac{m^{2}}{r^{2}}f_{+}{\cal E}-g_{+}{\cal
E}\right],
\end{equation}
where
\begin{subequations}
\begin{eqnarray}
\label{fplus} f_{+} &=&\frac{2\alpha^{2}+r(\alpha^{2})^{\prime}}
{\omega_{c}^{2}+4\alpha^{2}-(\Omega-m\omega_{+})^{2}+r(\alpha^{2})^{\prime}}\\
\label{gplus} g_{+}
&=&\left(\frac{m}{r}\right)\frac{1}{\Omega-m\omega_{+}}\frac{d}{dr}
\left[\sqrt{\omega_{c}^{2}+4\alpha^{2}}f_{+}\right].
\end{eqnarray}
\end{subequations}

For the angular frequency
$\omega_{-}=\frac{1}{2}[-\omega_{c}-\sqrt{\omega_{c}^{2}+4\alpha^{2}}]$,
an exactly parallel calculation yields $f_{-}$ and $g_{-}$. This
result shows an interesting symmetry that $f_{+}\rightarrow f_{-}$
by means of the substitution $\omega_{+}\rightarrow \omega_{-}$ or
vice versa. It should stressed that this calculation for $f_{-}$
and $g_{-}$ should be carried out to make sure our algebras are
indeed correct, although it is somewhat tedious.

Since the electron motion in the $r - \theta$ plane has slow and
fast rotations, and the generalized momentum (canonical momentum)
is the average of the mechanical momenta in the presence of a
static magnetic field, we take $f=[f_{+}+f_{-}]/2$ which is given
by

\begin{equation}
\label{ffactor} f =
[2\alpha^{2}+r(\alpha^{2})^{\prime}]\frac{\Theta^{2}}{\Theta^{4}
  -m^{2}(\omega_{c}^{2}+4\alpha^{2})(\Omega+m\omega_{c}/2)^{2}},
\end{equation}
where $\Theta^{2}=(1-m^{2}/4)(\omega_{c}^{2}+4\alpha^{2})-
(\Omega+m\omega_{c}/2)^{2} +r(\alpha^{2})^{\prime}$.

It should be stressed that an argument similar to that of Weinberg
\cite{Weinberg67} for taking an average based on the probability
of rotation in either (positive or negative) direction may not
hold in the presence of an external field. Since the equation of
motion can be written as $d{\bm
p}/dt=e[\bm{E}+\bm{\beta}\times\bm{B}]$, it is indeed correct to
take the canonical momentum in our analysis. This clearly shows
the inadequacy of Weinberg's model. And yet in the limit of
$\bm{B}^{ext}_{0}\rightarrow 0$, our final results reduce to those
of Weinberg \cite{Weinberg67}.

Similarly, we take $g=[g_{+}+g_{-}]/2$ and then is given by
\begin{eqnarray}
\label{gfactor} g
&=&-\left(\frac{m^{2}}{r}\right)\frac{(\Omega+m\omega_{c}/2)^{2}}
     {(\Omega^{2}-m^{2}\alpha^{2}+m\omega_{c}\Omega)}
     \left[\frac{(\omega_{c}^{2}+4\alpha^{2})f(r)}
     {\Theta^{2}}\right]^{\prime}\nonumber\\
  & & +\frac{1}{2}\left(\frac{m^{2}}{r}\right)
     \frac{(\omega_{c}^{2}+4\alpha^{2})^{1/2}}
     {\Omega^{2}-m^{2}\alpha^{2}+m\omega_{c}\Omega}
     \left[(\omega_{c}^{2}+4\alpha^{2})^{1/2}f(r)\right]^{\prime}
\end{eqnarray}

It should be noticed that, for a uniform beam,
$(\alpha^{2})^{\prime}=0$ which simplifies algebras immensely.
Henceforth we will assume that a uniform beam is coasting in the
over-dense plasmas and that the beam is completely neutralized.
For the special case of a uniform beam density profile, it is
possible to obtain a wave equation similar to that of Weinberg
\cite{Weinberg67}.

By taking $n(r)=n\theta(r-a)$, it follows from
Eqs.~(\ref{ffactor}) and (\ref{gfactor})

\begin{equation}
       f(r)=\begin{cases}
          1-\eta^{2} & \text{if $r<a$},\\
          0          & \text{if $r>a$}.
          \end{cases}
\end{equation}
Here $\eta^{2}$ is defined as
\begin{equation}
\label{Eta1}
\eta^{2}=1-\frac{2[(4-m^{2})+(1-m^{2}/4)\lambda_{c}^{2}-\bar{\lambda}^{2}]}
           {[(4-m^{2})+(1-m^{2}/4)\lambda_{c}^{2}-\bar{\lambda^{2}}]^{2}
           -m^{2}(4+\lambda_{c}^{2})\bar{\lambda}^{2}}
\end{equation}

Similarly,
\begin{equation}
g(r)=-\zeta^{2}\frac{\delta(r-a)}{a},
\end{equation}
where
\begin{equation}
\label{Zeta1}
\zeta^{2}=2m^{2}(1-\eta^{2})\frac{1+\lambda_{c}^{2}/4}
          {\bar{\lambda}^{2}-m^{2}(1+\lambda_{c}^{2}/4)}
          \left[1+\frac{2\bar{\lambda}^{2}}
          {\bar{\lambda}^{2}-(1-m^{2}/4)\lambda_{c}^{2}-(4-m^{2})}\right],
\end{equation}
with $\bar{\lambda}=\lambda+m\lambda_{c}/2$,
$\lambda=\Omega/\alpha$, and $\lambda_{c}=\omega_{c}/\alpha$. Here
$\alpha^{2}=\omega_{\beta}^{2}=2\pi e^{2}n\beta^{2}/(\gamma M)$
for a uniform beam density.

Substitution of these expressions for the perturbed density in
Eq.~(\ref{wave1}) yields

\begin{equation}
\label{wave2} \frac{1}{r}\frac{d}{dr} \left(r[1-f(r)]\frac{d{\cal
E}}{dr}\right) -\frac{m^{2}}{r^{2}}[1-f(r)]{\cal E} +
[q^{2}+g(r)]{\cal E}=0.
\end{equation}
 The detailed algebra leading to this wave equation is
 straightforward, but it is somewhat tedious. Similar to that of
 a two-dimensional vibrating membrane problem, the solution of
 Eq.~(\ref{wave2}), which is finite and satisfies the boundary
 conditions for a finite plasma channel \cite{Weinberg67}, is given
 by Hankel functions \cite{Jeffreys78}.
 \begin{equation}
 \label{efield}
 {\cal E}(r)=\begin{cases}
          J_{m}(qr/\eta) & \text{if $r<a$},\\
          H^{1}_{m}(qr)-\alpha_{m}J_{m}(qr) & \text{if $r>a$}.
          \end{cases}
 \end{equation}
Here $H^{1}_{m}(qr)=J_{m}(qr)+iN_{m}(qr)$

${\cal E}(r)$ by Eq.~(\ref{efield}), together with the boundary
conditions at the surface of the plasma channel of radius $b$
specifies the eigenvalue $\alpha_{m}$,
\begin{equation}
\label{eigenvalue} \alpha_{m}=\begin{cases}
          H^{(1)}_{m-1}(qb)/J_{m-1}(qb) & \text{if $m>1$}\\
          H^{(1)}_{1}(qb)/J_{1}(qb)     & \text{if $m=0$}.
          \end{cases}
\end{equation}
Integrating Eq.~(\ref{wave2}) over $a-\varepsilon$ and
$a+\varepsilon$, we obtain
\begin{equation}
\label{jump1}
 a{\cal E}^{\prime}(a+\varepsilon) -a\eta^{2}{\cal
E}^{\prime}(a-\varepsilon)=\zeta^{2}{\cal E}(a),
\end{equation}
Substitution of appropriate solutions of the wave equation from
Eq.~(\ref{efield}) into the left hand side of Eq.~(\ref{jump1})
and rearrangement of terms in the limit $\varepsilon \rightarrow
0$ yields the dispersion relation:
\begin{equation}
\label{dispersion1}
\eta\left(\frac{J^{\prime}_{m}(qa/\eta)}{J_{m}(qa/\eta)}\right)+
\frac{\zeta^{2}}{qa}=\frac{[H^{(1)}_{m}(qa)]^{\prime}-
\alpha_{m}(qb)J^{\prime}_{m}(qa)}{H^{(1)}_{m}(qa)-
\alpha_{m}(qb)J_{m}(qa)},
\end{equation}
where $m \geq 0$.

 This dispersion relation is identical in form to that
of Weinberg \cite{Weinberg67}, but differs in $\eta$ and $\zeta$.
In the limit of vanishing $B^{ext}$, the dispersion relation goes
over into that of Weinberg \cite{Weinberg67}. But it should be
noticed that the boundary condition at the edge of the plasma
column in Weinberg's analysis \cite{Weinberg67} is not valid,
since there is no external magnetic field that confines the plasma
channel in experiments. It is therefore apparent that his
dispersion relation Eq.~(1.9) [and Eq.~(12)] is inconsistent with
the problem he has posed in his paper and is in
self-contradiction.

\section{\label{sec:level1} Analysis of the Dispersion Relation for a
Uniform Beam}

We shall be interested here mainly in the resistive hose ($m=1$)
instability in the low and high frequency limits since the
resistive hose mode is the one that leads to the loss of a beam.
We note from Eq.~(\ref{dispersion1}) that, since $q^{2}=4\pi
i\sigma\omega/c^{2}$, it may be possible to obtain a rather simple
solution which holds to a higher order of approximation in the low
and high frequency limits. The method of making such an
approximation lies in the realization that the conductivity of the
plasma remains fixed for a given plasma density and the
approximation of Bessel functions in the asymptotic limits as a
function of $\omega$ readily available.

Thus the analytic solutions of Eq.~(\ref{dispersion1}) can be
studied in low and high frequency regimes with the asymptotic
limits of Bessel functions; the two asymptotic solutions should
then be connected smoothly by analytic continuation. For each $m$,
we may classify the modes as in Weinberg \cite{Weinberg67}. This
classification is not entirely trivial, since the external
magnetic field introduces new modes by removing the degeneracy
found in Weinberg's analysis \cite{Weinberg67}. The central
question in determining the efficiency of beam transport by means
of plasma focusing is to find which instability would affect the
beam transport most significantly. We present here only an outline
of the classification similar to that of Weinberg
\cite{Weinberg67} with emphasis on the resistive hose mode
($|\omega| \ll\sigma$) that affects the beam transport most
dangerously, since if the $m=1$ hose mode occurs, the entire beam
can be lost. Moreover the two-stream mode for which $\omega \sim
\sigma$ has been already treated in detail \cite{Frieman62}.

Returning to Eq.~(\ref{dispersion1}) we expand the left-hand side
(LHS) and the right-hand side (RHS), using the following
identities in Bessel functions:
\begin{subequations}
\label{identities}
 \begin{eqnarray*}
 \label{Bessel}
 H^{(1)}_{m}(qa) &=& J_{m}(qa)+iY_{m}(qa),\\
 J_{m}^{\prime}(qa) &=& (1/q)[qJ_{m-1}(qa)-(m/a)J_{m}(qa)],\\
 H_{m}^{(1)\prime}(qa) &=&
 (1/q)[qH_{m-1}^{(1)}(qa)-(m/a)H^{(1)}_{m}(qa)],
 \end{eqnarray*}
 \end{subequations}
as
\begin{subequations}
\begin{eqnarray}
\label{Left}
 LHS &=&
\eta\left[\frac{J_{m-1}(qa/\eta)}{J_{m}(qa/\eta)}-m(\frac{\eta}{qa})\right]\\
\label{Right}
 RHS &=& -\frac{m}{qa}
+\frac{Y_{m-1}(qa)J_{m-1}(qb)-
Y_{m-1}(qb)J_{m-1}(qa)}{Y_{m}(qa)J_{m-1}(qb)-Y_{m-1}(qb)J_{m}(qa)}.
\end{eqnarray}
\end{subequations}

With Eqs.~(\ref{Left}) and (\ref{Right}), Eq.~(\ref{dispersion1})
becomes
\begin{eqnarray}
\label{dispersion2}
\eta\left[\frac{J_{m-1}(qa/\eta)}{J_{m}(qa/\eta)}\right] &=&
-\frac{m}{qa}(1-\eta^{2})-\frac{\zeta^{2}}{qa} \nonumber \\
& &-\frac{Y_{m-1}(qa)J_{m-1}(qb)-
Y_{m-1}(qb)J_{m-1}(qa)}{Y_{m}(qa)J_{m-1}(qb)-Y_{m-1}(qb)J_{m}(qa)}.
\end{eqnarray}
Here $\eta^{2}$ and $\zeta^{2}$ are defined by Eq.~(\ref{Eta1})
and Eq.~(\ref{Zeta1}).

\subsection{\label{sec:level2} Low Frequency Regime: $|q|\rightarrow 0$}

For $|qa|\ll 1$, the dispersion relation Eq.~(\ref{dispersion1})
can be rewritten as

\begin{equation}
\label{lowfreq1}
\frac{\eta}{qa}\left(\frac{J_{m}^{\prime}(qa/\eta)}
                {J_{m}(qa/\eta)}\right)=
                -\frac{(m+\zeta^{2})}{(qa)^{2}} + {\cal O}(1),
\end{equation}
where $m\neq 0$, and

\begin{equation}
\label{lowfreq2}
\frac{\eta}{qa}\left(\frac{J_{m}^{\prime}(qa/\eta)}
                {J_{m}(qa/\eta)}\right)=
               \frac{1}{2}\left(b^{2}/a^{2}-1\right)+{\cal
               O}(q^{2}),
\end{equation}
where $m=0$, $qb\ll 1$.

 Here we have used the identity
 $H^{(1)}_{m}(qa)=J_{m}(qa)+iY_{m}(qa)$. We now classify the modes
 into those for which $|qa/\eta|\rightarrow 0$ at low  frequency
 and designate $C$ and $D$ modes, otherwise we call $A$ and $B$
 modes.

For $m=1$, $qa\ll 1$, and $qb \ll 1$ in Eq.~(\ref{dispersion2}),
we obtain the following expansion of the left hand-side of  the
equation:
\begin{equation*}
\frac{\eta}{qa}\frac{J_{1}^{\prime}(qa/\eta)}{J_{1}(qa/\eta)}
\rightarrow \frac{\eta^{2}}{(qa)^{2}}
\left[1-\frac{(qa)^{2}}{(2\eta)^{2}}-
\frac{q^{4}a^{4}}{96\eta^{4}}- \cdots - \right],
\end{equation*}
and similarly the right hand-side can be expanded as
\begin{eqnarray*}
-\frac{1}{qa}+\frac{Y_{0}(qa)J_{0}(qb)-Y_{0}(qb)J_{0}(qa)}
              {Y_{1}(qa)J_{0}(qb)-Y_{0}(qb)J_{1}(qa)}\\
              & \approx&  \frac{1}{qa}\left[1-(qa)^{2}\ln(b/a)
              +\frac{1}{2}(qa)^{4}\ln^{2}(b/a)+\frac{1}{2}\ln(b/a)\right]\\
              &  & -\frac{1}{qa} \left[\frac{1}{4}(qa)^{4}(b^{2}-a^{2}) + \cdots
              \right],
\end{eqnarray*}

Next we rewrite Eq.~(\ref{Eta1}) as
\begin{eqnarray}
\label{Eta2} \eta^{2}& =
&1-\frac{[3(1+\lambda_{c}^{2}/4)-\bar{\lambda}^{2}]}
         {[\bar{\lambda}^{2}-3(1+\lambda_{c}^{2}/4)]^{2}-
         4(1+\lambda_{c}^{2}/4)\bar{\lambda}^{2}}\\
        & \approx & \frac{1}{3}+\frac{1}{6}\lambda_{c}^{2}
         -\frac{14}{27}(1-\lambda^{2}/2+ \cdots +
         )\bar{\lambda}^{2} + \cdots + ,
\end{eqnarray}
and similarly rewrite Eq.~(\ref{Zeta1})
\begin{eqnarray}
\label{Zeta2}
 \zeta^{2} & = &
\frac{2(1+\lambda_{c}^{2}/4)}{\bar{\lambda}^{2}-(1+\lambda_{C}^{2}/4)}
          \left[1-\frac{2\bar{\lambda}^{2}}{3(1+\lambda_{c}^{2}/4)
          -\bar{\lambda}^{2}}\right](1-\eta^{2})\\
        & \approx &  -\frac{4}{3}(1-\lambda_{c}^{2}/4)-\frac{4}{3}
        (1-\lambda_{c}^{2}/4)
        \left(\frac{10}{9}\bar{\sigma}^{2} -\frac{5}{9}\bar{\sigma}^{4}+
       \cdots + \right),
\end{eqnarray}
where $\bar{\sigma}^{2}=\bar{\lambda}^{2}/(1+\lambda_{c}^{2}/4)$.

As shown by Weinberg \cite{Weinberg67}, for $|qa/\eta| \ll 1$  we
substitute the expressions Eq.~(\ref{Eta2}) and Eq.~(\ref{Zeta2})
to obtain an implicit dispersion relations
\begin{eqnarray}
\bar{\lambda}^{2}+\bar{\lambda}^{4}+\cdots +  & = &
\frac{1}{4}\lambda_{c}^{2}(1-\lambda_{c}^{2}/2)-\frac{1}{2}(1-\lambda_{c}^{2}/2)
\{\frac{1}{4}(1-\lambda_{c}^{2}/4)+\ln(b/a)\}(qa)^{2} \nonumber \\
& + & \frac{1}{4}(1-\lambda_{c}^{2})\{\frac{7}{16}+
\frac{1}{32}\lambda_{c}^{2}(1-\lambda_{c}^{2})+
\ln^{2}(b/a)+\ln(b/a)\}(qa)^{4} \nonumber  \\
& - & \frac{1}{8}(1-\lambda_{c}^{2})q^{4}a^{2}b^{2} + \cdots +.
\end{eqnarray}

Here we have assumed $|qa|\ll 1$ and $|qb|\ll 1$. In the limit
$q^{2}\rightarrow 0$, $\bar{\lambda}$ is oscillatory, which is the
hose instability and shows that, in the presence of the external
field, the growth rate is reduced due to the restoring force by
the magnetic field. This is in agreement with the numerical result
although the effects are not significant \cite{Sobehart93}.

While it is possible to calculate the $\bar{\lambda}^{2}$ to the
order of $(qa)^{4}$ by an iteration technique as in Weinberg
\cite{Weinberg67}, we just write, for the sake of simplicity,
$\bar{\lambda}^{2}$ to the order of $q^{2}a^{2}$ instead:
\begin{equation}
\label{Hose1}
\lambda^{2}+\lambda_{c}\lambda=-\frac{1}{8}\lambda_{c}^{4}-
\frac{1}{2}(1-\lambda_{c}^{2}/2)\{\frac{1}{4}(1-\lambda_{c}^{2}/4)+
\ln(b/a)\}(qa)^{2}+ \cdots + .
\end{equation}
Here $\bar{\lambda}=\lambda+\lambda_{c}/2$,
  $\lambda=\pm(\omega-kv)/\omega_{\beta}$, and $q^{2}=(4\pi i
\sigma\omega)/c^{2}$, and
$\lambda_{c}^{2}=\omega_{c}^{2}/\omega_{\beta}^{2}$.

Thus,
\begin{equation}
\label{Hose2}
\lambda=\frac{1}{2}\left[-\lambda_{c}\pm(1-\lambda_{c}^{2})^{1/2}
\{\lambda_{c}^{2}-2[\frac{1}{4}(1-\lambda_{c}^{2}/4)+
\ln(b/a)]q^{2}a^{2}\}^{1/2}\right].
\end{equation}

Eq.~(\ref{Hose2}) is the dispersion relation for the hose
instability in the limit $q^{2}a^{2}\rightarrow 0$ and it shows
that, in the presence of the external magnetic field, the growth
rate is reduced due to the restoring force by the magnetic field
which is in agreement with numerical computations
\cite{Sobehart93}.

\subsection{\label{sec:level2}Unstable  Modes (A and B): $|qa/\eta|$ does not converge
to zero:} If $qa/\eta$ does not converge to zero, $\eta$ must then
converge to zero as fast as $|qa|$. Rewriting Eq.~(\ref{Eta1}) as
\begin{equation}
\label{Eta3} \eta^{2}=\frac{\Lambda^{2}
        -2\Lambda
        -m^{2}(4+\lambda_{c}^{2})\bar{\lambda}^{2}}
           {\Lambda^{2}-m^{2}(4+\lambda_{c}^{2})\bar{\lambda}^{2}},
\end{equation}

where
$\Lambda=(4-m^{2})+(1-m^{2}/4)\lambda_{c}^{2}-\bar{\lambda^{2}}$.

Hence the numerator of the above equation must be zero which
defines the modes of A and B modes:

\begin{equation}
\label{LambdaAB}
      \bar{\lambda}^{2}=\begin{cases}

          [(1+m^{2}/4)\lambda^{2}_{c}+(3+m^{2})]-
          \Delta_{m}^{1/2} & \text{(A mode)},\\

          [(1+m^{2}/4)\lambda^{2}_{c}+(3+m^{2})]+
          \Delta_{m}^{1/2} & \text{(B mode)},

          \end{cases}
\end{equation}
where
$\Delta_{m}=1+4m^{2}(1+\lambda^{2}_{c}/4)(3+\lambda^{2}_{c})$.

 With the external magnetic field, the degeneracy of $m=0$
and $m=2$ are removed in A and B modes found in Weinberg's
analysis \cite{Weinberg67}.

But as in Weinberg's analysis, $m+\zeta^{2}$ in the righthand side
of Eq.~(\ref{lowfreq1}) does not vanish for $m>0$ when
$\bar{\lambda}^{2}$ takes either value of the above A and B modes.

Hence we may carry out similar analysis iteratively for
$\bar{\lambda}^{2}$ from Eq.~(\ref{lowfreq1}) for $A_{mn}$ - mode
by the equation
\begin{equation}
\eta\rightarrow (qa)/j_{mn},
\end{equation}
where $j_{mn}$ is the n-th root of $J_{m}(x)=0$ which yields
\begin{equation}
\bar{\lambda}^{2}_{+}\rightarrow
(3+m^{2})+(1+m^{2})\lambda_{c}^{2}-\sqrt
\Delta_{1}-\frac{1}{2}\frac{(qa)^{2}}{j_{mn}^{2}}[2+\frac{m^{2}(4+\lambda_{c}^{2})-2}
{\sqrt\Delta_{1}}].
\end{equation}
Similarly, for $B_{mn}$ - mode we have
\begin{equation}
\bar{\lambda}^{2}_{-}\rightarrow
(3+m^{2})+(1+m^{2})\lambda_{c}^{2} +\sqrt
\Delta_{1}+\frac{1}{2}\frac{(qa)^{2}}{j_{mn}^{2}}
[2+\frac{m^{2}(4+\lambda_{c}^{2})-2}{\sqrt\Delta_{1}}],
\end{equation}
where $\Delta_{1}=m^{2}(4+\lambda_{c}^{2})(3+\lambda_{c}^{2})+1$.

Next for $m=0$, Eq.~(\ref{lowfreq2}) can be written as
\begin{equation}
\eta\rightarrow (qa)/y_{n}.
\end{equation}
Here $y_{n}$ is the n-th root of the equation
\begin{equation}
J_{0}^{\prime}(y)/(yJ_{0}(y))=\frac{1}{2}\left(b^{2}/a^{2}-1\right).
\end{equation}
Hence for $m=0$, there is no difference between the present
analysis and that of Weinberg's analysis except for the presence
$\lambda_{c}^{2}$ term which removes the degeneracy. Solving
Eq.~(\ref{Eta3}) by iteration, we obtain
\begin{equation}
\bar{\lambda}^{2}=\lambda_{c}^{2}+3 \pm 1-2(qa)^{2}/y_{n}^{2} +
\cdots.
\end{equation}

\subsection{\label{sec:level2} $ m=1$ Resistive Hose Instability: (C, D) modes}

Suppose $qa/\eta \rightarrow 0$. For $m \neq 0$, the left-hand
side of Eq.~(\ref{lowfreq1}) then becomes
\begin{equation}
\label{hose1}
\frac{\eta}{qa}\left(\frac{J_{m}^{\prime}(qa/\eta)}{J_{m}(qa/\eta)}\right)\approx
m \frac{\eta^{2}}{q^{2}a^{2}},
\end{equation}
with the aid of the identity
\begin{equation}
J_{m}^{\prime}(qa/\eta)=-J_{m+1}(qa/\eta)
+(m\eta/qa)J_{m}(qa/\eta).\nonumber
\end{equation}

And yet the condition that $qa/\eta \rightarrow 0$ implies that,
by Eq.~(\ref{Eta1}), $\bar\lambda$ does {\it not} take the
following value
\begin{equation}
\bar{\lambda}^{2}=(1+m^{2}/4)(4+\lambda^{2}_{c})-1\pm[m^{2}(4+\lambda^{2}_{c})
                 {(4+\lambda^{2}_{c})+1}]^{1/2},
\end{equation}
for which $\eta=0$.

With the condition $|qa/\eta|\rightarrow 0$ and with
Eq.~(\ref{hose1}), Eq.~(\ref{lowfreq1}) yields the following
condition:
\begin{equation}
\label{CandD}
 m\eta^{2}+ m+ \zeta^{2}=0,
\end{equation}
where $m>0$.

The zeros of this equation in which $\eta^{2}$ and $\zeta^{2}$ are
defined by Eq.~(\ref{Eta2}) and Eq.~(\ref{Zeta2}) defines the
modes of type C and D and is given by:
\begin{equation}
\label{CD} \bar{\lambda}^{2}=(m^{2}-2m +3/2)(1+\lambda_{c}^{2}/4)
\pm(1+\lambda^{2}_{c}/4)^{1/2}\left[\{3m(m-2)+9/4\}(1+\lambda^{2}_{c}/4)-
m(m-2)\right]^{1/2},
\end{equation}
where $\pm$ signs correspond to the modes of type C and D
respectively based on Weinberg's classification \cite{Weinberg67}.
Note also that if $\bar{\lambda}\rightarrow \lambda$ for
$\lambda_{c}=\omega_{c}/\alpha=0$, the results of Eq.~(\ref{CD})
in the limits go over B14 and B15 of Weinberg \cite{Weinberg67} as
they should.

\subsection{\label{sec:level2} High Frequency Regime: $|q|\rightarrow \infty$}

\subsubsection{\label{sec:level3} ${\cal A}$ Mode and ${\cal B}$ Mode}

For $|qa|\gg 1$ and $|qb|\gg 1$, with the aid of asymptotic
formula of the Bessel functions of the first kind $J_{m-1}(qa)$
and the second kind $Y_{m-1}(qa)$ for $qa \gg 1$,
\begin{equation}
Y_{m-1}(qa) \sim \left[\frac{2}{\pi
qa}\right]^{1/2}\cos[qa-\frac{\pi}{4}-\frac{1}{2}(m-1)\pi],
\end{equation}
and
\begin{equation}
J_{m-1}(qa) \sim \left[\frac{2}{\pi
qa}\right]^{1/2}\sin[qa-\frac{\pi}{4}-\frac{1}{2}(m-1)\pi],
\end{equation}

the dispersion relation Eq.~(\ref{dispersion2}) reduces to
\begin{equation}
\label{HFdispersion}
\eta\left(\frac{J^{\prime}_{m}(qa/\eta)}{J_{m}(qa/\eta)}\right)+
\frac{\zeta^{2}}{qa}\approx {\mathcal {O}(\frac{1}{qa})}.
\end{equation}

Next we divide the modes into those for which $|qa/\eta|$ remains
finite with $|q|\rightarrow\infty$ ($\cal{A}$ and $\cal{B}$ modes)
and those for which $|qa/\eta|\rightarrow\infty$ ($\cal{D}$ mode).

With some algebraic rearrangement, we may rewrite Eq.~(\ref{Eta1})
as
\begin{equation}
\label{Eta4}
\eta^{2}=1-\frac{2[(1-m^{2}/4)(4+\lambda_{c}^{2})-\bar{\lambda}^{2}]}
                {[\bar{\lambda}^{2}-(1+m/2)^{2}(4+\lambda_{c}^{2})]
                 [\bar{\lambda}^{2}-(1-m/2)^{2}(4+\lambda_{c}^{2})]}.
\end{equation}

It is easy to see from Eq.~(\ref{Eta3}) that $|qa/\eta|$ may
remain finite if the following conditions are met:
\begin{equation}
\label{HighfieldAB}
\bar{\lambda}^{2}_{\pm}\rightarrow\begin{cases}
          (1+m/2)^{2}(4+\lambda_{c}^{2}) & \text{$\cal {A}$ mode},\\
          (1-m/2)^{2}(4+\lambda_{c}^{2}) & \text{$\cal {B}$ mode}.
          \end{cases}
 \end{equation}

For a finite value of $|x|=|qa/\eta|$, we may rewrite
Eq.~(\ref{HFdispersion}) to obtain the $1/q^{2}$ term in
$\bar{\lambda}^{2}$,
\begin{equation}
\frac{J_{m}^{\prime}(x)}{xJ_{m}(x)}+\frac{\zeta^{2}}{(x
\eta)^{2}}\rightarrow 0,
\end{equation}
where $x=qa/\eta$.

The ${\cal A}$ and ${\cal B}$ modes for which
$\bar{\lambda}^{2}_{\pm}$ takes $(1 \pm
m/2)^{2}(4+\lambda_{c}^{2})$, and using those in Eq.~(\ref{Eta4})
and Eq.~(\ref{Zeta2}) we may write

\begin{equation}
\frac{\zeta^{2}}{\eta^{2}}=-\left(1-\frac{1}{\eta^{2}}\right)
\Gamma_{\pm}(m,\bar{\lambda}_{\pm}),
\end{equation}
where $\Gamma_{\pm}$ is defined as the following:
\begin{equation}
\Gamma_{\pm}(m,{\bar{\lambda}}_{\pm})=\frac{m^{2}}{2}\left(\frac{4(1+\lambda_{c}^{2}/4)}
               {{\bar
               {\lambda}}^{2}_{\pm}-m^{2}(1+\lambda_{c}^{2}/4)}\right)
              \left[1+\frac{{\bar\lambda}^{2}_{\pm}}{{\bar{\lambda}}^{2}_{\pm}
              -(1-m^{2}/4)\lambda_{c}^{2}-4(1-m^{2}/4)}\right].
\end{equation}

After a brief algebra, we obtain
\begin{equation}
\Gamma_{\pm}(m,{\bar{\lambda}}_{\pm})=\pm \frac{3}{4}
                                     \frac{(m \pm 2)}{(m \pm 1)}.
\end{equation}
Since $1/\eta^{2}\rightarrow 0$ as $\eta\rightarrow \infty$,
\begin{equation}
\frac{\zeta^{2}}{\eta^{2}}=\mp \frac{3}{4}
                                     \frac{(m \pm 2)}{(m \pm 1)},
\end{equation}
which is independent of $\lambda_{c}$. If
$\bar{\lambda}_{+}^{2}=(1+m/2)^{2}(4+\lambda_{c}^{2})$ for $\cal
{A}$ mode, then
\begin{equation}
\frac{\zeta^{2}}{\eta^{2}}=\frac{3}{4}
                                     \frac{(m + 2)}{(m + 1)}.
\end{equation}

Hence the dispersion relation for $\cal {A}$ mode is
\begin{equation}
\bar{\lambda}_{+}^{2}\rightarrow 4(1+m/2)^{2}(1+\lambda_{c}^{2}/4)
                     +\frac{3}{4}\frac{(m+2)}{(m+1)}\frac{x^{2}}{(qa)^{2}},
\end{equation}
where $x$ is the solution of the equation
\begin{equation}
xJ_{m}^{\prime}(x)=\pm \frac{3}{4} \frac{(2+m)}{(1+m)}J_{m}(x).
\end{equation}

When $\bar{\lambda}_{-}^{2}=(1-m/2)^{2}(4+\lambda_{c}^{2})$ for
$\cal {B}$ mode except for $m=1$ and $m=2$, we have

\begin{equation}
\frac{\zeta^{2}}{\eta^{2}}=\frac{3}{4}
                                     \frac{(m-2)}{(m -1)}.
\end{equation}

Hence the dispersion relation for $\cal {B}$ mode is
\begin{equation}
\bar{\lambda}_{+}^{2}\rightarrow 4(1-m/2)^{2}(1+\lambda_{c}^{2}/4)
                     +\frac{3}{4}\frac{(m-2)}{(m-1)}\frac{x^{2}}{(qa)^{2}},
\end{equation}
where $x$ is the solution of the equation
\begin{equation}
xJ_{m}^{\prime}(x)=\pm \frac{3}{4} \frac{(2-m)}{(1-m)}J_{m}(x),
\end{equation}
where both $m \neq 1$ and $m \neq 2$.

We also notice that our results do not go over to those of
Weinberg's analysis in the limit $\lambda_{c}\rightarrow 0$,
because there was an algebraic error in his analysis [see
equations (B26) and (B27)].

\subsubsection{\label{sec:level3} ${\cal D}$ Mode}

If $\eta^{2}$ remains finite but $\zeta^{2}\rightarrow \infty$,
then the mode $\cal {D}$ would take place in the beam. This
happens if $\bar{\lambda}^{2}\neq
(1\pm/2)^{2}(4\pm\lambda_{c}^{2})$ and
$\bar{\lambda}^{2}\rightarrow m^{2}(1+\lambda^{2}_{c})$ except for
$m=1$ for which the numerator of $\zeta^{2}$ becomes zero with
$\bar{\lambda}^{2}=m^{2}(1+\lambda_{c}^{2}/4)$. For $m=0$,
$\zeta^{2}$ is identically zero. Hence the type $\cal {D}$ mode
begins with $m \geq 2$.

In a high frequency limit, $|q|\rightarrow\infty$, $|qa|\gg 1$,
and $|qb|\gg 1$, $\cal {D}$ mode is given by
\begin{equation}
\bar{\lambda}^{2}=(1+\frac{\lambda_{c}^{2}}{4})+\frac{x^{2}}{2q^{2}a^{2}},
\end{equation}
where $x$ is a root of $xJ_{1}^{\prime}(x)=-3J_{1}(x)$ and
$qa/|\eta|\rightarrow \infty$.

\begin{figure}
\includegraphics[totalheight=0.45\textheight,viewport=100 50 400 400,height=4.5in,width=4.0in,keepaspectratio=true]{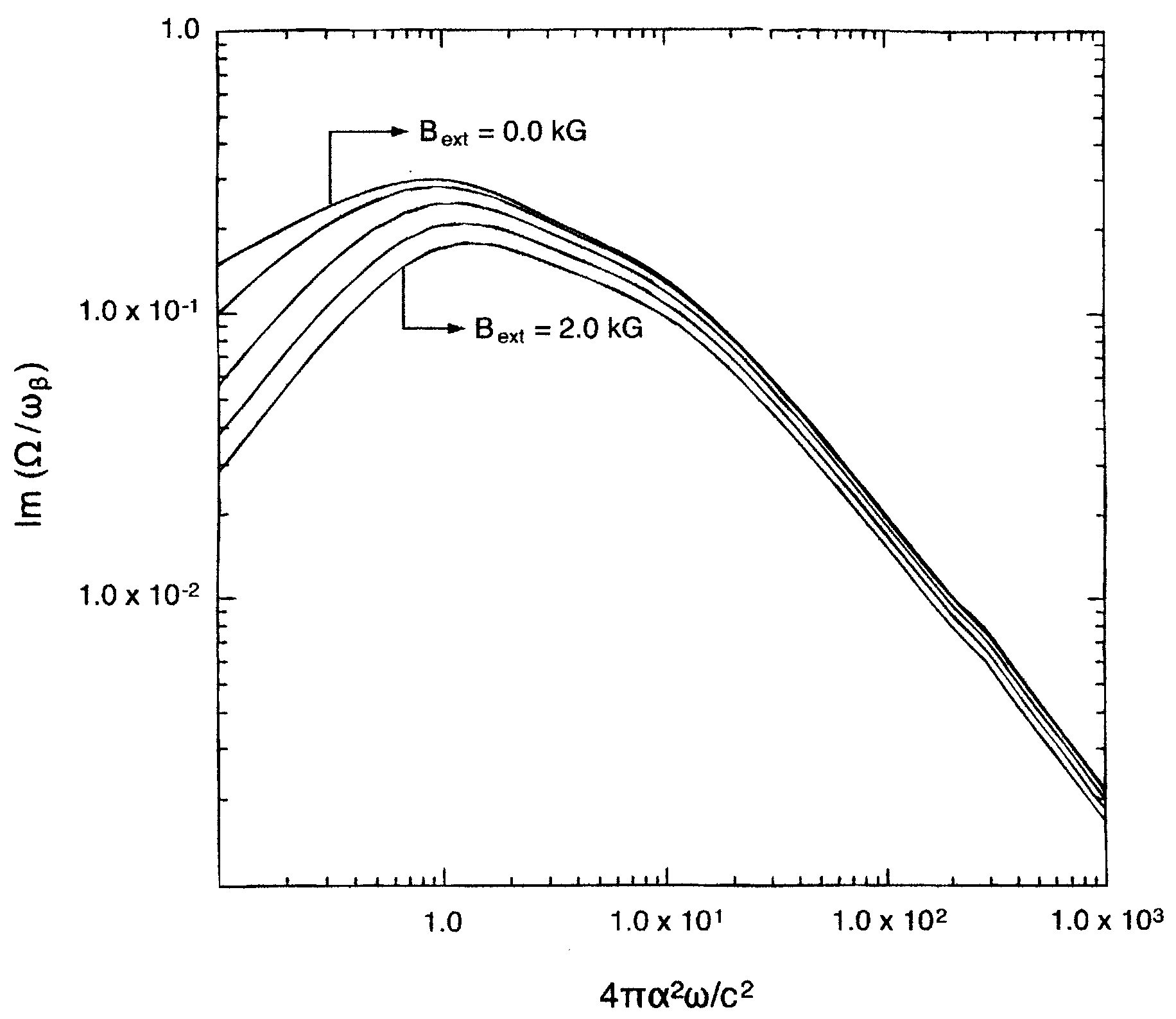}\vspace{2.0truecm}
\caption {The growth rate $Im (\Omega/\omega_{\beta})$ of the
hose-mode ($C_{1}$) as a function of frequency $4\pi
\sigma^{2}\omega/c^{2}$ for various external magnetic field
strengths $B_{i}$. The curve for $B_{ext}= 0.0 kG$ corresponds to
the Weinberg's solution for which the present model of plasma
channel breaks down. As the strength of the magnetic field is
increased by $\Delta B_{ext}=0.5 kG$, the growth rate decreases,
but not significantly.} \label{fig:fig2}
\end{figure}

We have carried out the above analysis to guide numerical work of
solving the dispersion relation for various instabilities,
Eq.~(\ref{dispersion2}), by a computer. The above analytical
results have been borne out in our detail numerical calculations.
We have found that the external magnetic field reduces the growth
rate somewhat, but not significantly since
$\omega_{c}=ecB^{ext}/{\cal E}$ in the relativistic domain. The
results (Figure \ref{fig:fig2}) are essentially same as the
Weinberg's analysis.

\section{\label{sec:level1} Discussion and Conclusion}

In conclusion we note that it may be somewhat confusing by
studying both a proton beam for a proton linac and an electron
beam for PHERMEX facility together, but the stability analysis
remains valid for both cases with a proper change of charge and
mass of a particle. The proposed technique of combination of the
synchrotron and the RFQ to accelerate a high-current beam to meet
the necessary energy requirements for the beam injection may
become feasible, since it takes so short a time to travel for the
beam in a short plasma channel that the resistive hose instability
may not develop to disrupt the integrity of the beam. Moreover,
the technique has been demonstrated in an electron accelerator by
Nakanishi group \cite{Nakanishi91}. However, the experimental
evidence we could find (including the plasma cone in a Be chamber
of PHERMEX facility) is too scant to permit any safe
generalization. Indeed it would be desirable to have experimental
demonstrations for a high-energy, high-current proton beam. Yet
the case we have discussed so far is conclusive enough to
demonstrate its feasibility. Without a proper reconditioning a
high-current beam by a dense plasma channel, a reliable operation
of a high-current, high-energy accelerator is highly unlikely.
However, the creation of a dense plasma-channel may complicate the
maintenance of a linac in a routine operation and may make it
difficult to perform any reliable experiments. Still a potential
difficulty in developing a reliable beam injector for a
high-current, high-energy accelerator remains the major stumbling
block for the the large hadron collider (LHC) at CERN or APT
project at LANL. In any practical, realistic sense, the LHC is
doomed to failure \cite{Ptoday07,NPR07,Ptoday08}, but particle
physicists could still turn to astrophysical observations
\cite{Lindley93}.

\newpage

\end{document}